\documentclass[11pt]{article} 
\usepackage{hyperref} 
\pdfoutput=1 
 
\begin{document} 
 
\title{Ignition sequence of an annular multi-injector combustor} 
 
\author{M.~Philip, M.~Boileau, R.~Vicquelin, T.~Schmitt, \\ D.~Durox, J.-F.~Bourgoin, S.~Candel\\
\\\vspace{6pt} Ecole Centrale Paris, Laboratoire EM2C, CNRS UPR 288\\
Grande Voie des Vignes, 92295 Ch\^atenay-Malabry, France} 
 
\maketitle 
 
 
\begin{abstract} 
Ignition is a critical process in combustion systems. In aeronautical combustors, altitude relight capacities are required in case of accidental extinction of the chamber. A simultaneous study of light-round ignition in an annular multi-injector combustor has been performed on the experimental and numerical sides. This effort allows a unique comparison to assess the reliability of Large-Eddy Simulation (LES) in such a configuration. Results are presented in fluid dynamics videos. 
\end{abstract} 
 
 
\section{Video description} 
 
After describing the annular multi-injector combustor, the LES ignition sequence is first presented, then compared to the synchronized experimental visualization. Very good agreement is observed and further analysis of the LES results allows to clearly identify the different stages of the light-round process.\\

\par
The annular combustor is composed of 16 swirl injectors fed with a propane/air mixture of equivalence ratio 0.76, providing a total thermal power of 52 kW in nominal conditions. The bulk velocity in the injection pipe is 17.1 m/s and each injector is characterized by a swirl number of 0.8. The experimental visualization shows that successful ignition of all burners is achieved within 50 ms.\\

\par
The LES is performed with the AVBP solver. The mesh is composed of 310 millions tetrahedra and extends from the plenum feeding lines upstream of the combustion chamber till the exterior atmosphere. The WALE subgrid stress model is used and propagation of the turbulent premixed flame front is described by the FTACLES model.\\

\par
For further information on the experimental setup, please see \cite{Bourgouin20131398}.

\bibliographystyle{plain}

\end{document}